\pdfoutput=1
\documentclass[twocolumn,superscriptaddress,amsmath,amssymb,aps,prb]{revtex4-2}
\usepackage{amsmath}
\usepackage{amssymb}
\usepackage{graphicx}
\usepackage{color}
\usepackage{booktabs}
\usepackage[colorlinks=true,linkcolor=blue,citecolor=blue]{hyperref}
\usepackage{multirow}
\usepackage[table,xcdraw]{xcolor}

\newcommand{\be}{\begin{equation}}
\newcommand{\ee}{\end{equation}}
\newcommand{\bea}{\begin{eqnarray}}
\newcommand{\eea}{\end{eqnarray}}
\newcommand{\bse}{\begin{subequations}}
\newcommand{\ese}{\end{subequations}}

\newcommand{\Tn}{$T_{\textrm N}$}

\usepackage{siunitx}
\newcommand{\bfS}{{\bf S}}

\newcommand{\redbf}[1]{}

\newcommand{\memome}[1]{}
\newcommand{\req}[1]{Eq.~(\ref{#1})}

\begin{document}

\title{Spin Dynamics in the Antiferromagnetic Oxypnictides and Fluoropnictides: LaMnAsO, LaMnSbO, and BaMnAsF}

\author{Farhan Islam}
\author{Elijah Gordon}
\author{Pinaki Das}
\author{Yong Liu}

\author{Liqin Ke}

\affiliation{Ames Laboratory, Ames, Iowa 50011, USA}

\author{Douglas L. Abernathy}
\affiliation{Quantum Condensed Matter Division, Oak Ridge National Laboratory, Oak Ridge, Tennessee 37831, USA}

\author{Robert J. McQueeney}
\author{David Vaknin}
\affiliation{Ames Laboratory, Ames, Iowa 50011, USA}
\affiliation{Department of Physics and Astronomy, Iowa State University, Ames, Iowa 50011, USA}

\date{\today}

\begin{abstract}
Inelastic neutron scattering (INS) from polycrystalline antiferromagnetic LaMnAsO, LaMnSbO, and BaMnAsF are analyzed using a $J_1-J_2-J_c$ Heisenberg model in the framework of the linear spin-wave theory.  All three systems show  clear evidence that the nearest- and next-nearest-neighbor interactions within the Mn square lattice layer ($J_1$ and $J_2$) are both antiferromagnetic (AFM). However, for all compounds studied the competing interactions have a ratio of $2J_2/J_1 < 1$, which favors the square lattice checkerboard AFM structure over the stripe AFM structure. The inter-plane coupling $J_c$ in all three systems is on the order of $\sim 3\times10^{-4}J_1$, rendering the magnetic properties of these systems with quasi-two-dimensional character. The substitution of Sb for As significantly lowers the in-plane exchange coupling, which is also reflected in the decrease of the N{\'e}el  temperature, $T_{\rm N}$.  Although BaMnAsF shares the same MnAs sheets as LaMnAsO, their  $J_1$ and $J_2$ values are substantially different. Using density functional theory, we calculate exchange parameters $J_{ij}$ to rationalize the differences among these systems.
\end{abstract}
\maketitle
\section{Introduction}

Manganese (Mn) based pnictide compounds with Mn$Pn$ ($Pn =$ P, As, Sb, and Bi)  layers have  been in the spotlight by virtue of their intriguing magnetic properties, most notably the recently discovered Dirac semimetals $A$Mn$Pn_2$ ($A=$ Ca, Sr, and Ba)~\cite{Park2011,Farhan2014,Huang2017}. The quasi two-dimensional (2D) $A$Mn$Pn_2$  have been recognized as the three-dimensional (3D) analogs of the 2D graphene with linearly dispersing bands that cross at the Fermi energy~\cite{Farhan2014}.  Generally, the Mn atoms are arranged in a square lattice or in a slightly distorted  orthorhombic lattice and  undergo antiferromagnetic (AFM) ordering. Of particular interest is the coupling of magnetism to Dirac fermions, which in ideal cases can deliver a Weyl semimetal with unique bulk magnetotransport and optical properties. A few compounds exhibit  uniform canting with a finite ferromagnetic (FM)  moment  that can further remove the degeneracy of the Dirac bands to furnish Weyl states~\cite{Park2011,Wang2011,Sanchez-Barriga2016,Liu2017,Rahn2017,Liu2019}.

$A$Mn$_2Pn_2$  ($A=$ Ca, Sr, Ba) is another   class of AFM semiconductors with similar Mn$Pn$ layers and with partially localized Mn moments.  These compounds are not known to possess Dirac-like bands, but  can become metallic with doping~\cite{An2009,Pandey2012a,Lamsal2013}.  It has been reported that the substitution of K for Ba in (Ba$_{1-x}$K$_x$)Mn$_2$As$_2$ shows a novel magnetic ground state below $T_{\rm C} \simeq 100$ K, in which itinerant ferromagnetism  associated with the As bands  coexists with a collinear local-moment AFM ordering associated with the Mn atoms with $T_{\rm N}\simeq  480$ K (for $x =0.2$)~\cite{Pandey2013,Ueland2015}. We note that other reports associate the FM in this system to simple canting of the  AFM magnetic moments that gives rise to the observed weak-FM signal~\cite{Glasbrenner2014}.  

It is clear that systematic studies of the evolution of the magnetism among all of these
square-lattice Mn pnictides are necessary. The magnetism in these systems is dominated by the Mn$Pn$ square layer where Mn-$Pn$-Mn  superexchange couplings  between the nearest- and next-nearest-neighbor (NN and NNN, $J_1-J_2$) Mn spins lead to a checkerboard-type AFM order. Easy axis anisotropy results in Mn moments that point normal to the square layers. The AFM Mn$Pn$ planes are coupled via intervening layers by a much weaker AFM or FM exchange coupling $J_c$.  Analysis of magnetic excitations obtained by inelastic neutron scattering (INS)  of polycrystalline (Ba$_{1-x}$K$_x$)Mn$_2$As$_2$ determined  competing AFM $J_1-J_2$ exchange interactions and the much weaker interplane coupling $J_c$ of a Heisenberg model~\cite{Ramazanoglu2017}. 
In fact,   INS studies of  other layered Mn$Pn$ compounds, such as BaMn$_2$Bi$_2$ \cite{Calder2014}, and  the topological semimentals $A{\mathrm{MnBi}}_{2}$ ($A=\mathrm{Sr}, \mathrm{Ca}$)\cite{Rahn2017} and  ${\mathrm{YbMnBi}}_{2}$.\cite{Soh2019}, have also established the presence of competing AFM  $J_1-J_2$ exchange couplings.

\begin{figure}[h]
\includegraphics[width=1.5in]{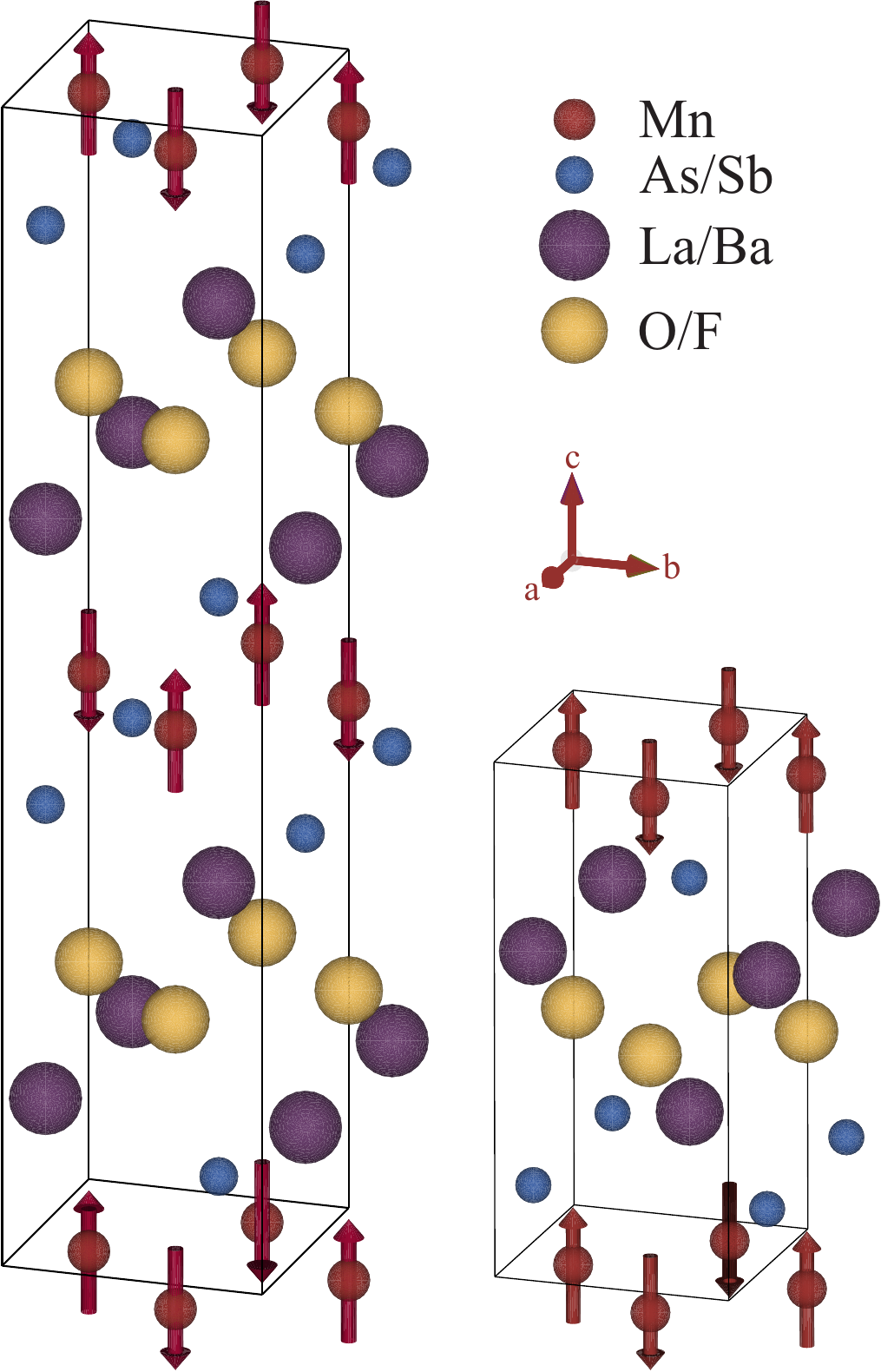}
\caption{(Color online) Crystal and magnetic structure of LaMnSbO and LaMnAsO (C-type) on the right and BaMnAsF (G-type) on the left.}
\label{Fig:Crystal}
\end{figure}

 Another  class of  AFM compounds that shares similar Mn$Pn$ planes is $R$Mn$PnB$ ($R=$ La, Ce, Pr, Ba, ...and $B =$ O and F, referred to here as Mn-1111 compounds)~\cite{Emery2010,Kimber2010,Tsukamoto2011,Emery2011,Lee2012,Zhang2015, McGuire2016,Saparov2013,Zhang2016}. Recently it has been suggested that one such compound, LaMnAsO, can be hole doped by substitution of Sr for La and undergo insulator-to-metal transition and exhibit thermoelectric properties~\cite{Sun2012,Hanna2013}.

In this manuscript, we report on the measurements and analysis of INS data from polycrystalline samples of Mn-based 1111 pnictides --- LaMnSbO, LaMnAsO, and BaMnAsF. They all belong to the $P4/nmm$ space group, and both LaMnSbO and LaMnAsO have C-type AFM order, whereas BaMnAsF has G-type AFM order, as depicted in Fig.\ \ref{Fig:Crystal}. By analyzing the spin-waves in the framework of a Heisenberg model, we determine the exchange interactions  $J_1$, $J_2$, and $J_c$ and show that  all compounds demonstrate a significant competitive AFM NNN interaction ($J_2$), that places these systems close to a magnetic instability between checkerboard and stripe AFM order. The very weak inter-plane interaction ($J_c$) renders the spectra with quasi-2D characteristics. Density functional theory (DFT) calculations confirm the magnetic ground states, the average magnetic moment, and the exchange parameters determined experimentally.  Confirmation of these energy scales provides theoretical grounds for designing new materials for potentially novel ground states in Mn$Pn$ systems, such as spin liquids or magnetic topological materials.

\section{Experimental Details}
\textit{Sample Preparation}: Polycrystalline samples of LaMnAsO, LaMnSbO, and BaMnAsF were synthesized by a solid-state reaction method. The stoichiometric chemicals of La and Ba pieces, Mn, As, Sb, MnO and BaF$_2$ powder were weighed and mixed in a glovebox under argon atmosphere. The mixtures were pressed into pellets under a pressure of 12 MPa. The pellets were loaded into alumina crucibles and sealed in quartz tubes. The quartz ampoules were slowly heated up to 500 $^\circ$C at a ramping rate of 100 $^\circ$C/h. After a dwell time of 6 hours, the ampoules were heated up to 780 $^\circ$C/h at the same rate and held at that temperature for 6 hours. These prereacted samples were crushed and ground in the glovebox. The powder was pressed into pellets and sintered at 1100 1 $^\circ$C/h for 12 hours in an evacuated quartz tube. After sintering, the furnace was  cooled down to room temperature at a rate of 200  $^\circ$C/h. To improve the homogeneity and get rid of impurity phases, the final step was repeated once.

\begin{figure}[!hb]
\includegraphics[width=3.1 in]{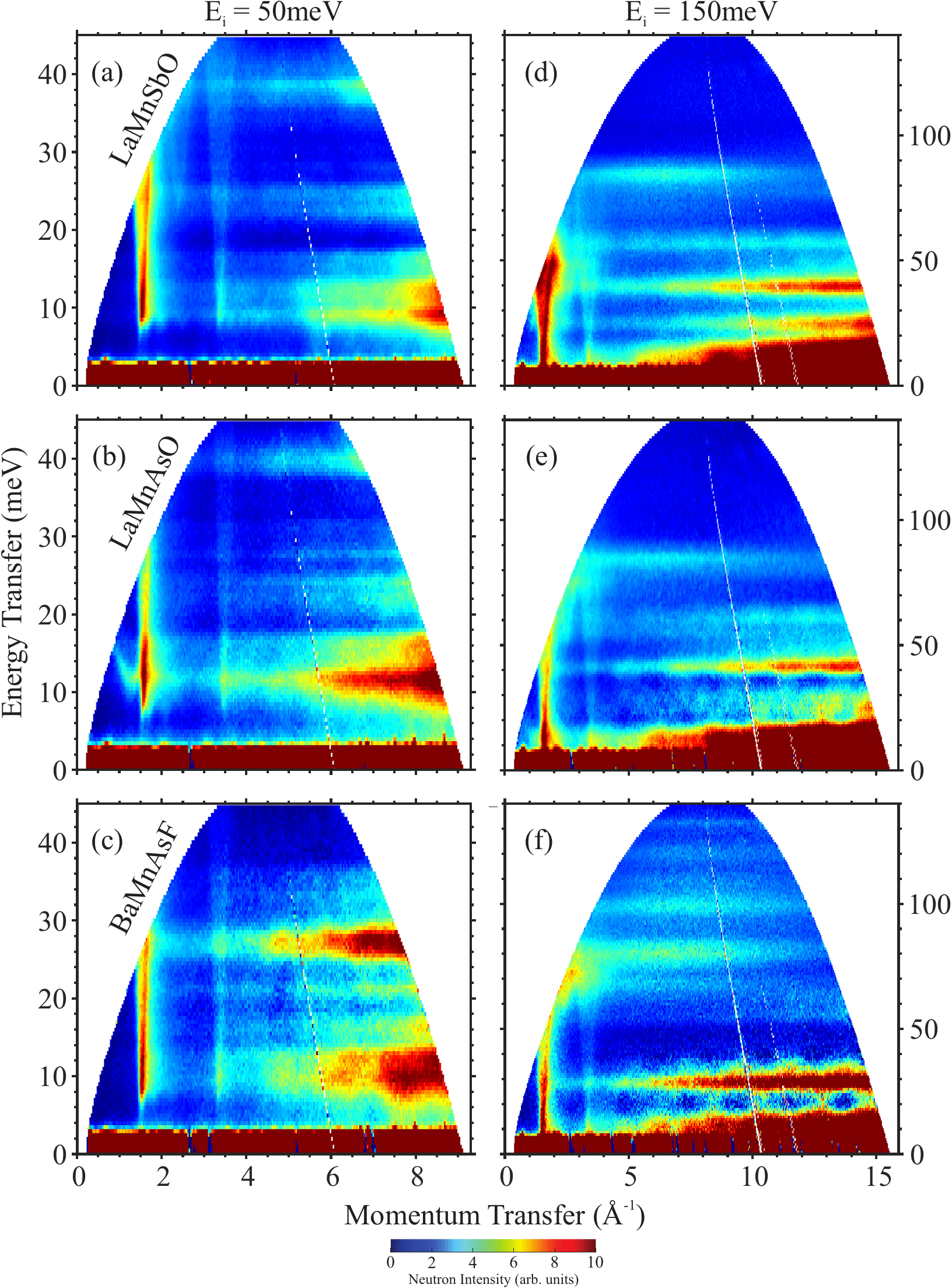}
\caption{INS intensities $S(Q,E)$ for $E_i =50$ and 150 meV  for (a) and (d) LaMnSbO, (b) and (e) LaMnAsO, and (c) and (f) BaMnAsF, respectively as indicated. Columns of scattering emanating at $\sim 1.6 $  and at 3.5 {\AA}$^{-1}$ are due to magnons.  }
\label{Fig:rawdata}
\end{figure}
Powder x-ray diffraction (XRD) measurements were performed on a PANalytical MPD diffractometer using Co K$\alpha$ radiation. Magnetization measurements were performed by using Physical Property Measurement System (PPMS, Quantum Design) equipped with Vibrating Sample Magnetometer (VSM). 
All three 1111 compounds  crystallize in a tetragonal $P4/nmm $  crystal symmetry, with the lattice parameters listed in Table\ \ref{tab:params} , with no change in crystal symmetry down  to base temperature ($T = 12$~K). LaMnSbO and LaMnAsO adopt a C-type AFM ground state and BaMnAsF into  a G-type with varying {\Tn} as listed in Table~\ref{tab:params} (based on Refs.~[ 
\onlinecite{Emery2010,Kimber2010,Tsukamoto2011,Emery2011,Lee2012,Zhang2015, McGuire2016,Saparov2013,Zhang2016}]).

\textit{Inelastic Neutron Scattering}: INS measurements were carried out on the ARCS spectrometer at the Spallation Neutron Source at Oak Ridge National Laboratory.
Each polycrystalline sample was placed in a cylindrical aluminum sample can and mounted on the cold tip of a closed-cycle helium cryostat.
Measurements were performed at $T = 10$ K with incident energies, $E_i = 50, 150, 300,$ and $500$ meV with an energy resolution of $3-5$~\% of $E_i$.
The data were corrected for both aluminum (sample holder) and hydrogen scattering (due to surface adsorption of water by exposure of the polycrystalline sample to air. See more details in SM \cite{SM}).
Incoherent nuclear scattering from a vanadium standard was used to correct for the variation of the detector efficiency.
The dynamical structure factor $S(Q,E)$, where $Q$ is the momentum transfer and $E$ is the energy transfer, were used to get $Q$- and $E$-cuts for refined fitting.

\textit{Modeling with $\textsc{SpinW}$}:
We use $\textsc{SpinW}$, a \textsc{matlab} library, to model the magnetic excitations and fit the INS data~\citep{Toth2015}.
We set up the crystal properties for each compound using documented lattice constants, space-group, atomic position of magnetic atoms, neutron scattering form factor, and magnetic structure.
We specify the Heisenberg interactions between $ab$-plane nearest neighbor ($J_1$) and next nearest neighbor ($J_2$), c-axis nearest neighbor ($J_c$), and single-ion anisotropy ($D$). The powder-averaged spin wave spectrum are calculated  by averaging over a large number of momentum transfer vectors on the surface of a sphere of radius $Q$. 
The Heisenberg spin Hamiltonian for the $J_1$-$J_2$-$J_c$-$D$ model can be written as:

\begin{equation}
\begin{split}  
  H = &J_{1}\sum_{i\neq j\in ab}{\bfS_i\cdot\bfS_j} + J_{2}\sum_{i\neq k\in ab}{\bfS_i\cdot\bfS_k} \\
      & + J_{c}\sum_{i\neq l\in c}{\bfS_i\cdot\bfS_l}  + D\sum_{i}{(S_i^z)^2}.    
\label{eq:Heisenberg}      
  \end{split}
\end{equation}
We compare the exchange parameters and single-ion anisotropy values extracted from experiments with those obtained from DFT.

\textit{DFT calculational details:}
Spin-polarized DFT+$U$ calculations, within the Dudarev scheme~\cite{Dudarev1998}, were carried out in the Vienna Ab-initio Simulation Package (\textsc{vasp})~\cite{Kresse1999,Kresse1996} by employing the projected-augmented wave method~\cite{Blochl1994}.
The exchange-correlation functional used is the generalized gradient approximation of Perdew, Burke, and Ernzerhof~\cite{Perdew1996}.
The difference between the effective on-site Coulomb and exchange parameters, denoted as $U$ (0--5 eV), was used to simulate additional Mn $d$-orbital on-site electron-electron correlations.
Plane wave cutoff energy was set at 500 eV and the energy threshold for calculation was set at $10^{-6}$ eV.
Exchange parameters are calculated using an energy-mapping analysis~\cite{Xiang2013}.
The total energies of four different collinear spin configurations are calculated and mapped to ~\req{eq:Heisenberg} to extract the three exchange parameters, $J_1$, $J_2$, and $J_c$ (computational details can be found in the Supplemental Material\cite{SM}).

To determine the single-ion anisotropy term, $SD$ in ~\req{eq:Heisenberg}, we calculate the magnetocrystalline anisotropy energy (MAE) of each compound. MAE originates from the spin-orbit coupling (SOC)~\cite{ke2015prb,ke2019prb}. We include SOC using the second-variation method~\cite{koelling1977jpcs,Li1990,Shick1997} in our calculations.
Starting from the experimental spin configuration of each compound, we calculate $SD$=$E_{a}{-}E_{c}$, where $E_{a}$ and $E_{c}$ are the total energies (per Mn) of the system with spins aligned along the $a$ or $c$ axis, respectively, and $S$ is the magnitude of Mn spin.

\section{Results and Discussion}
\begin{table}
\caption{Lattice parameters $a$ and $c$ of LaMnAsO, LaMnSbO, and BaMnAsF in space group $P4/nmm$. The atomic positions of La and Ba,  are at ($\frac{1}{4},\frac{1}{4},z_A$)  and As and Sb at ($\frac{1}{4},\frac{1}{4},z_P$). $J_1$, $J_2$, $J_c$, and $D$ are the exchange couplings between intralayer NN, NNN, interlayer NN, and the single-ion anisotropy, respectively as obtained from our modeling of INS data. $\Delta$ is the energy gap.}
\renewcommand{\arraystretch}{1.0}
\setlength{\tabcolsep}{0.9em}
\begin{tabular}{llll}
\hline
\hline
         		& LaMnSbO 	& LaMnAsO 		& BaMnAsF 	\\ \hline
$a$  ({\AA})     & 4.236    	& 4.111   		& 4.26    	\\
$c$    ({\AA})   & 9.545    	& 9.026   		& 9.559   	\\
$z_A$        	& 0.619    	& 0.633   		& 0.661   	\\
$z_{P}$	 		& 0.181    	& 0.168   		& 0.154   	\\
$T_{\rm N}$ (K)	& 255 		& 360			& 338		\\
\hline
$SJ_{1}$	(meV)	& 40(4)		& 48(4)			& 35(4)		\\
$SJ_{2}$	(meV)	& 17(2)		& 18(3)			& 10(2)		\\	
$SJ_{c}$	(meV)	& -0.01$^*$	& -0.01$^*$		& 0.01$^*$	\\	
$SD$	(meV)	& -0.07(2)	& -0.045(30)		& -0.06(4)	\\	
$J_2/J_1$		& 0.42(6)	& 0.38(7)		& 0.29(6)	\\
$\Delta$ (meV)	& 8(2)		& 9(2)			& 7(2)		\\
\hline \hline
\label{tab:params}
\end{tabular}
*The value for $J_c$ is the upper limit modeling is not sensitive to values in the range of 0.01 to $10^{-4}$ meV. Numbers in bracket are the uncertainty in the last digit of a value. 
\end{table}
\begin{figure}[h]
\includegraphics[width=3.1in]{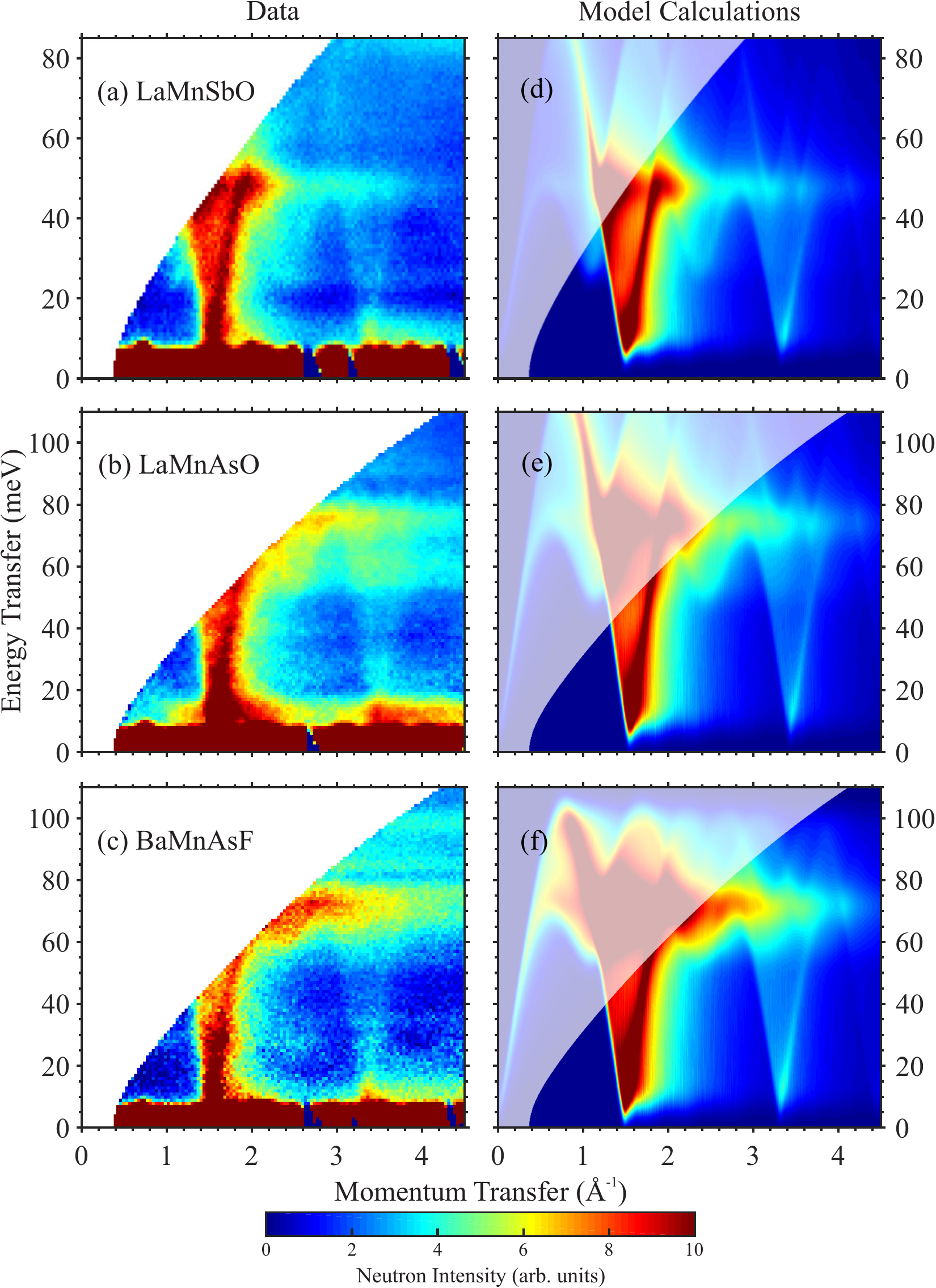}
\caption{(Left column) measured inelastic neutron scattering data at $E_i$ = 150 meV for (a) LaMnSbO, (b) LaMnAsO, and (c) BaMnAsF as indicated. (Right column,) (d-f) Corresponding calculated spectra using the best fit parameters given in  Table\ \ref{tab:params}. The shaded areas in the calculated panels are kinematically inaccessible regions for neutrons at the specified energy and set up.}
\label{Fig:ins150mev}
\end{figure}

\begin{figure}[!hb]
\includegraphics[width=3.1in]{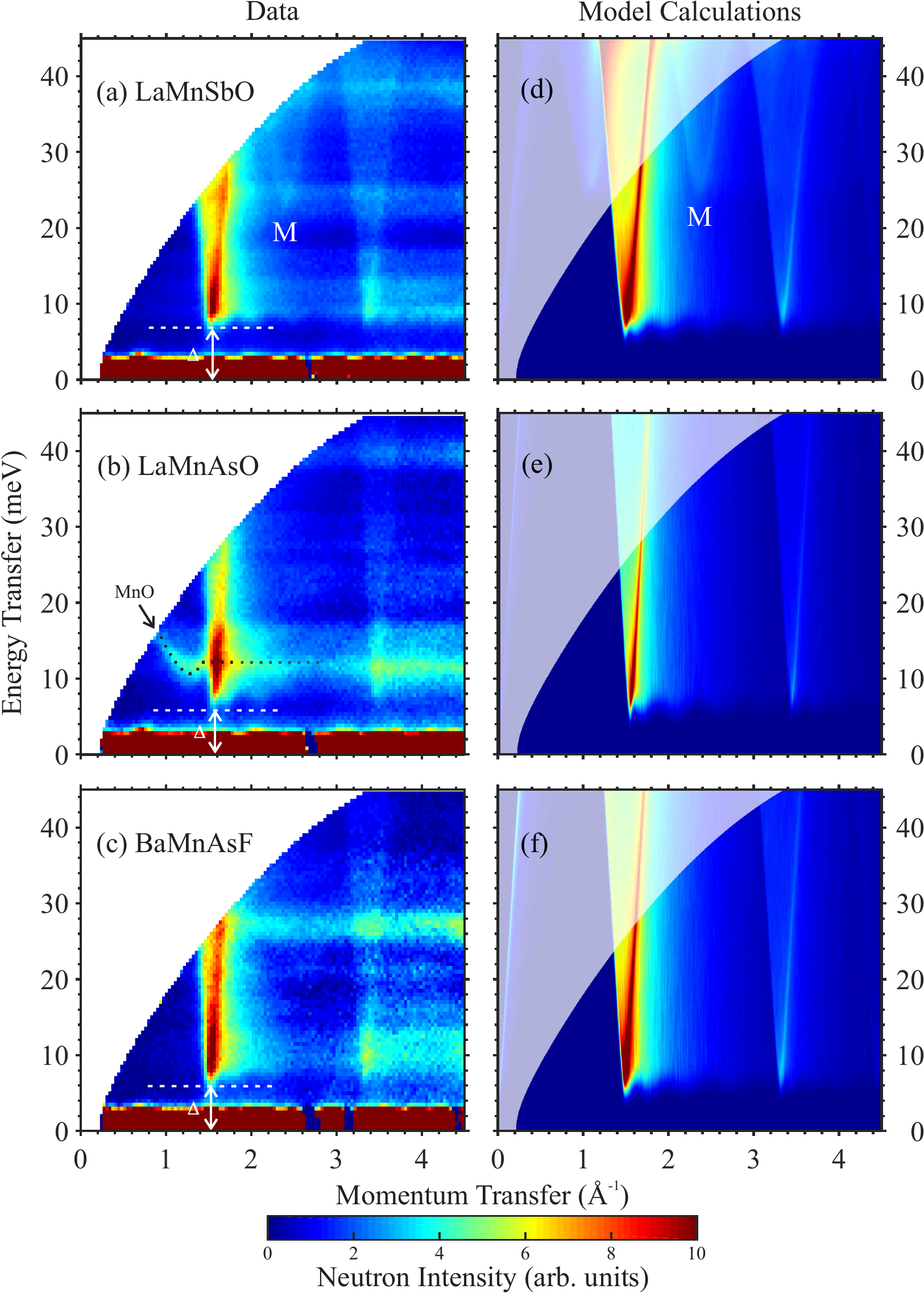}
\caption{ (Left column) measured inelastic neutron scattering data at $E_i$ = 50 meV for (a) LaMnSbO, (b) LaMnAsO, and (c) BaMnAsF as indicated. (Right column) (d-f) Corresponding calculated spectra using the best fit parameters given in  Table\ \ref{tab:params}. The shaded areas in the calculated panels are kinematically inaccessible regions neutrons at the specified energy and set up. Notice the weak  but detectable minimum at the M-point for LaMnSbO that also shows in the calculations.}
\label{Fig:ins50mevgap}
\end{figure}
 
\subsection{Measured and Simulated Spin-waves Spectra}
Figure \ref{Fig:rawdata} shows  INS intensity maps, proportional to $S(Q,E)$ for polycrystalline LaMnSbO, LaMnAsO, and BaMnAsF at $T=10$ K  for two incident energies $E_i = 50$ and 150 meV. Each $S(Q,E)$ map has a major contribution in the elastic region  near  $E=0$  due to elastic Bragg reflections and incoherent scattering (neutron energy loss is positive). The $S(Q,E)$ data also includes strong intensities that grow as $Q^2$ due to INS from phonons. The magnetic INS in our samples form steep columns, that emanate from (1 0 0) and (1 2 0) magnetic Bragg reflections for LaMnSbO and LaMnAsO, and from the (1 0 $\frac{1}{2}$) and (1 2 $\frac{1}{2}$) reflections for BaMnAsF that slightly open into cones at high energies. Due to the fast-falling off of the magnetic form factor of Mn$^{2+}$, magnon scattering intensity practically vanishes for  $Q \geq 4.5$ {\AA}$^{-1}$.   To analyze the magnetic spectra, we focus our analysis to $Q \leq 4.5$ {\AA}$^{-1}$. In this region, the intensity due to magnetic scattering is still contaminated by phonon scattering and other background contributions that can complicate the modeling.  We cleaned up the phonon signal by fitting phonon peaks in the high-$Q$ region $Q \geq 4.5$ {\AA}$^{-1}$ with a Gaussian function and estimating its intensity in the low-Q region by interpolation from the high-Q region.
 
To model the magnetic spectra we follow a procedure similar to that provided in Ref.~[\onlinecite{Ramazanoglu2017}] and using the Heisenberg Hamiltonian in Eq.\ (\ref{eq:Heisenberg}).
 In the linear approximation,  spin operators in Eq.\ \eqref{eq:Heisenberg} are transformed into bosonic operators with the Holstein-Primakoff approximation, leading to  spin wave dispersion relations
\begin{equation}
\begin{aligned}
\bigg[ \frac{\hbar \omega ( \bf{q})}{2S} \bigg]^2 =
&\bigg [ 2J_1 - J_2(2-\cos{q_xa}-\cos{q_ya}) \\
&- J_c(1 - \cos{q_zc})  + D \bigg]^2 \\
&-\bigg [ J_1 \{ \cos{\frac{(q_x + q_y)a}{2}} + \cos{\frac{(q_x - q_y)a}{2}} \}\bigg]^2
\end{aligned}
\label{eq:3}
\end{equation}
for C-type structure, and
\begin{equation}
\begin{aligned}
\bigg[ \frac{\hbar \omega ( \bf{q})}{2S} \bigg]^2 =
&\bigg [ 2J_1 - J_2(2-\cos{q_xa}-\cos{q_ya}) + J_c + D \bigg]^2 \\
&-\bigg [ J_1 \{ \cos{\frac{(q_x + q_y)a}{2}} + \cos{\frac{(q_x - q_y)a}{2}} \} \\ 
& +J_c\cos{(\frac{q_zc}{2})}\bigg]^2
\end{aligned}
\label{eq:4}
\end{equation}
for G-type structure where $\textbf{q}$ is the wave vector measured relative to a $\Gamma$-point at a magnetic Bragg peak, and $a$ and $c$ are the lattice parameters for the tetragonal $P4/nmm $ unit cell.

We first make a rough estimate of $J_1$ and $J_2$ and subsequently refine $D$ and $J_c$ by fitting to the low energy portion of the magnetic spectrum.  After refining $D$ and $J_c$, we perform fits to the full magnetic spectrum by fixing $D$ and $J_c$ and varying $J_1$ and $J_2$.  This process is repeated until good convergence is achieved, although additional constraints, described below, were necessary to optimize $J_1$ and $J_2$.

Using $\textsc{SpinW}$ we calculate magnon dispersion and the powder-averaged intensities $S(Q,E)$  by Monte Carlo sampling of 50000 $Q$-vectors for a given magnitude of $Q$, from 0.1 - 4.2 {\AA}$^{-1}$ as shown in Figs.\ \ref{Fig:ins150mev} and \ref{Fig:ins50mevgap}. Different $E-$ and $Q$-cuts were fit by using the non-linear least-squares process to capture major features of the INS spectra. 

\begin{figure}[h]
\includegraphics[width= 2.5in]{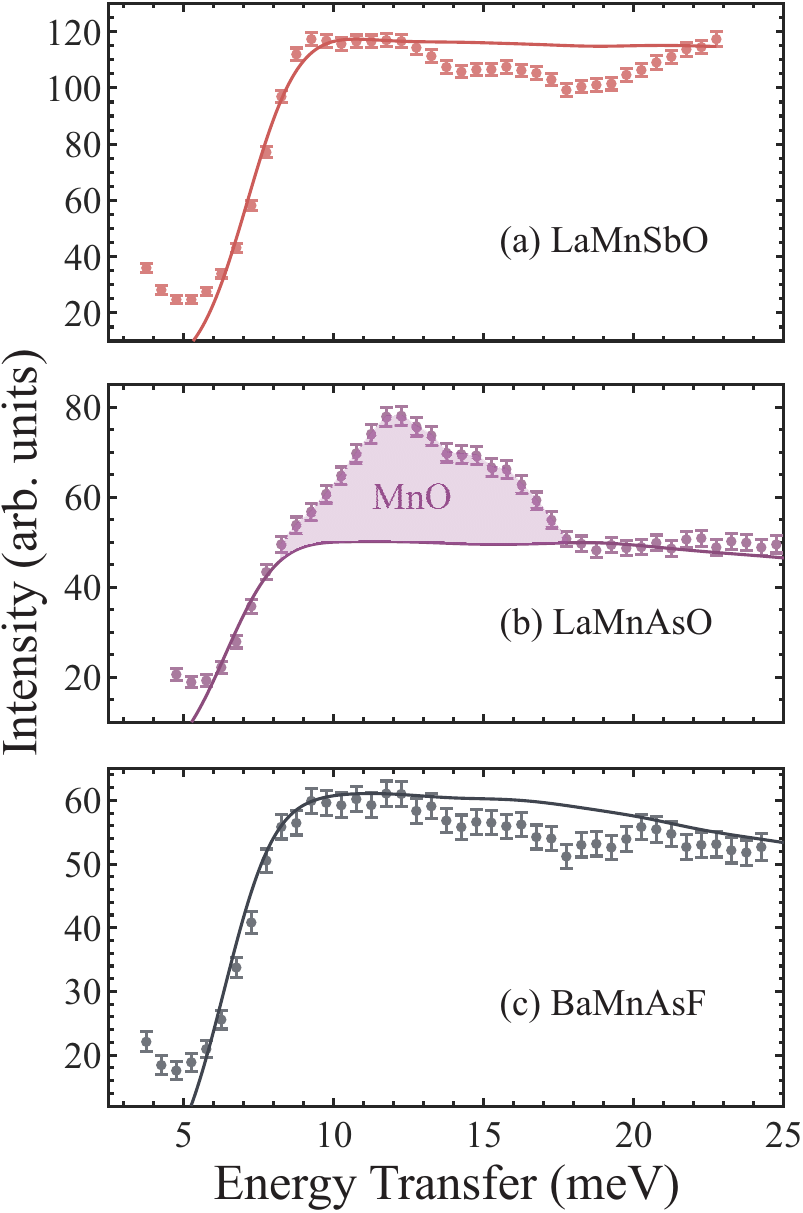}
\caption{Energy cuts at constant-Q around $Q_{(100)} \pm0.2$ \AA$^{-1}$ for (a) LaMnAsO and (b) LaMnSbO and at the  $Q_{(10\frac{1}{2})} \pm0.2 $ \AA$^{-1}$ for (c) BaMnAsF.  The solid lines are best fit to the experimental data using the parameters in Table \ \ref{tab:params}. For LaMnAsO we identify significant magnetic INS contribution from  MnO that is present as an impurity phase.}
\label{Fig:gapcut}
\end{figure}

\begin{figure}[h]
\includegraphics[width=2.5in]{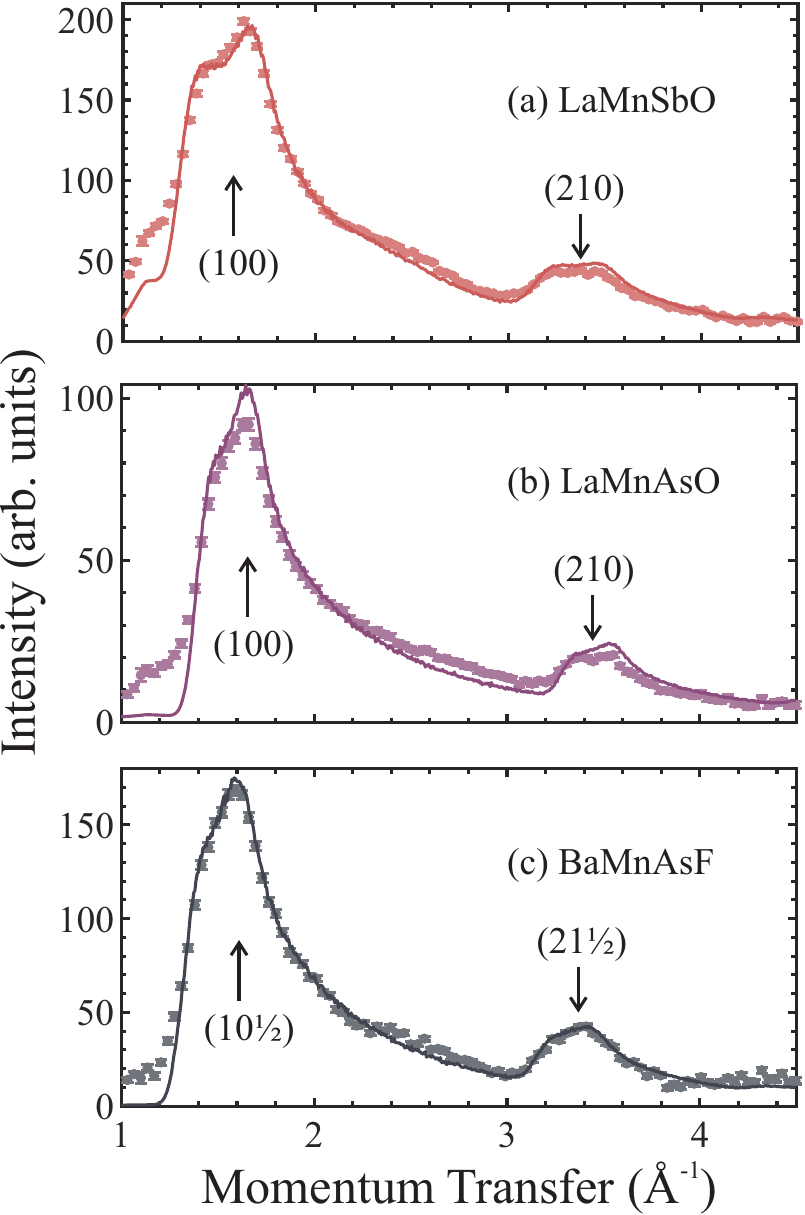}
\caption{Q-cut around low energy region ($E\approx 25 \pm 5$ meV) for (a) LaMnSbO, (b) LaMnAsO, and (c) BaMnAsF showing the 2D nature of the spin excitations. The region is chosen on the $E_i=150 $ meV data with relatively cleaner region where signal from phonons is absent. These cuts are used to estimate an upper limit for $J_c$. }
\label{Fig:jccut}
\end{figure}

{\it Spin gap ($\Delta$) and single ion anisotropy $D$:} To estimate $D$  we focus on Fig.\  \ref{Fig:ins50mevgap} with spectra obtained at $E_i = 50$ meV, where it can be seen that there is a gap in the spin-wave spectrum of $\approx 6$ meV for each compound. Figure\ \ref{Fig:gapcut} shows energy averaged over a limited range of $Q$ and centered at $Q_{(100)} \pm0.2$ \AA$^{-1}$ for LaMnAsO and LaMnSbO and at the  $Q_{(10\frac{1}{2})} \pm0.2$ \AA$^{-1}$ for BaMnAsF, obtained from data shown in Fig.\ \ref{Fig:ins50mevgap}.  The solid lines are best fit to the experimental data using the parameters listed in Table\ \ref{tab:params}. For LaMnAsO we identify significant magnetic INS contribution from MnO that is present as an impurity phase (for details on the magnetic INS contribution of MnO polycrystalline see Ref.~[\onlinecite{Ramazanoglu2017}]).

\begin{figure}[h]
\includegraphics[width=3in]{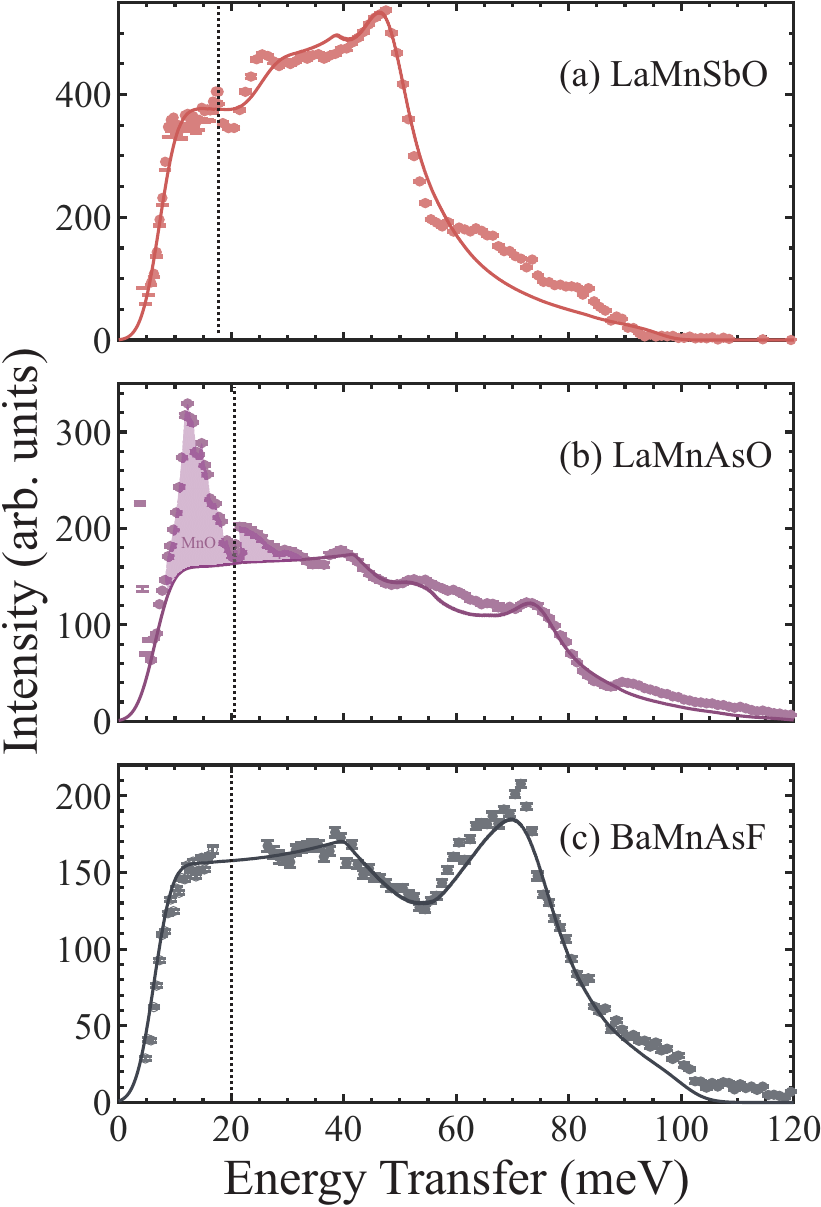}
\caption{Energy-cuts for the full $S(Q,E)$ spectrum for (a) LaMnSbO, (b) LaMnAsO, and (c) BaMnAsF. Additional phonons and other backgrounds were subtracted by similar methods described in \cite{Ramazanoglu2017}. Circular symbols on the left and right side of the vertical dashed line denote the data extracted from $E_i$ = 50 meV and 150 meV respectively. (b) We determined the presence of MnO in LaMnAsO sample and the magnon signal from AFM MnO is shown in the shaded region. (c) Obvious phonon signal was detected near 20 meV of BaMnAsF spectrum which could not be subtracted in a systematic manner. Hence we decided to omit those points.}
\label{Fig:vanhovecut}
\end{figure}

{\it Two-dimensionality of the systems:} The $J_c$ term in Eqs. (\ref{eq:3}) and  (\ref{eq:4}) determines the interlayer correlations. For all three samples, we find that the value of $J_c$ is negligibly small. Although we kept the value of $|J_c|$ fixed at 0.01 meV,  this serves as an upper bound as modeling the data using $|J_c|$  as small as $10^{-4}$ meV yields similar results.
As  $|J_c|$ increases above 0.01 meV, we visually notice  that columns of excitations emanate from (10$L$) magnetic Bragg peaks in our models, which is not observed experimentally. To get more insight into $J_c$, we make $Q$ cuts near roughly $E\simeq 25$ meV where the INS data is  cleaner and free from phonon and multiple scattering signals as shown in Fig. \ref{Fig:jccut}.  The   $Q$-cuts in Fig. \ref{Fig:jccut} all show characteristic quasi-2D features with a tail that extends to large $Q$ values. This is similar to a Warren lineshape which corresponds to the powder averaging of rod of scattering. This behavior can be contrasted with  similar cuts in the INS of the more 3D-like BaMn$_2$As$_2$ for which the scattering is modulated with peaks that are near ($H$0$L$) reflections~\cite{Ramazanoglu2017}. 

\begin{figure}
\includegraphics[width=3.3in]{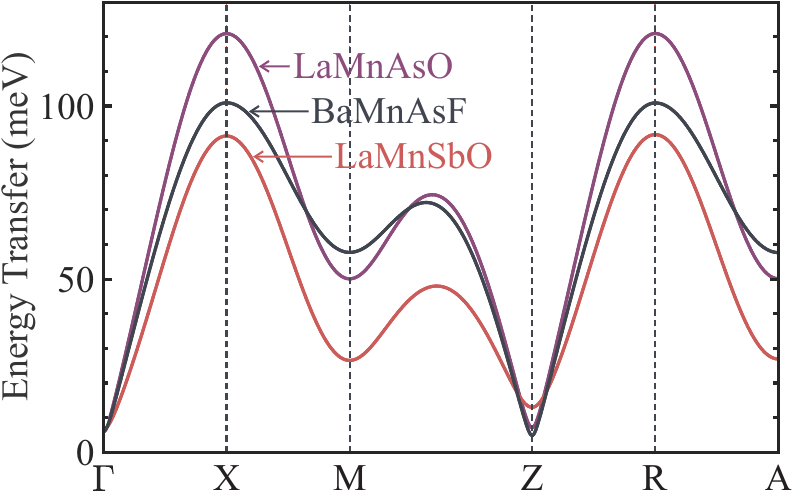}
\caption{Calculated spin-waves dispersions  along principal directions of  single crystals using the best fit parameters obtained in this study.}
\label{Fig:singlecrystal}
\end{figure}

{\it Relation between $J_1$ and $J_2$ and their determination}:
Fixing $J_c$ and $D$, we proceed by systematically varying $J_1$ and $J_2$ to model  energy-cuts as shown in Fig.\ \ref{Fig:vanhovecut}. We calculate  $\chi^2$ values for numerous combinations  of $J_1$ and $J_2$ to search for its minimum to obtain the best fit to the data. A  color map of $\chi^{2}(J_1,J_2)$ is shown in Fig. S1 and the optimal values are listed in Table\ \ref{tab:params} (note that in Fig. S1 we present 1/$\chi^{2}(J_1,J_2)$ for color enhancement purposes).  

Figure S1 shows that the minima in $\chi^2$ form a shallow valley which does not allow for a precise determination of $J_1$ and $J_2$.  We can improve this situation by exploiting the extrema (van Hove singularities) in the spin wave dispersion to further constrain the values. For example, Fig.\ \ref{Fig:singlecrystal} shows the spin wave dispersion obtained for each compound using the parameters from Table I.  The maximum between the M- and Z-point gives rise to a van-Hove singularity that results in a peak in the magnetic spectra.  Whereas the M-point energy is evident in the measured and calculated spectra for LaMnSbO, see Fig.\ \ref{Fig:ins50mevgap} (a) and (d),  for LaMnAsO and BaMnAsF we can only estimate this point from $E_i=150$ meV with larger uncertainty. Our best estimates of the minimum at the M-point  is at 23, 50, and 58 meV with a standard deviation of 2, 5, and 5 meV for LaMnSbO, LaMnAsO, and BaMnAsF, respectively. Similarly, looking at $E_{i} = 150$ and $E_i=300$ meV data, (see, Fig. S2 in SM) we estimate that the spin-wave bandwidth (corresponding to the  X-point) of LaMnSbO, LaMnAsO, and BaMnAsF to be at $90\pm3$, $120\pm3$, and $95\pm5$ meV. It is worth noting that the kinematic constraint of neutron does not allow us to get a good handle on the X-point, which would have significantly narrowed the uncertainties  of $J_1$ and $J_2$.

\renewcommand{\arraystretch}{1.0}
\setlength{\tabcolsep}{0.9em}
\begin{table*}[!ht]
\caption{List of exchange parameters $SJ_i$, single-ion anisotropy $SD_c$, in-plane Mn-Mn NN distance $d_{\rm NN}$, and distance between adjacent Mn$Pn$ layers in various systems. }
\begin{tabular}{lllllll}
\hline
\hline
         						& $SJ_{1}$ (meV)	& $SJ_{2}$ (meV)	& $SJ_{\rm c}$ (meV)	& $SD$ (meV)	& $d_{\rm NN}$ ({\AA})	& $d_{\rm c}$ ({\AA})	\\ \hline
LaMnSbO						    	& 40(4)    		& 17(2)   		& -0.01    		& -0.07(2)		& 2.99				& 9.54				\\
LaMnAsO							& 48(4)    		& 18(3)   		& -0.01   		& -0.045(30)		& 2.91				& 9.03				\\
BaMnAsF							& 35(4)    		& 10(2)   		& 0.01   		& -0.29(6)		& 3.01				& 9.60				\\
\hline
SrMnBi$_{2}$ \citep{Rahn2017}		& 21.3(2)   		& 6.39(15)  		& 0.11(2)   		& -0.31(2)		& 3.24				& 5.78				\\
CaMnBi$_{2}$ \citep{Rahn2017}		& 23.4(6) 		& 7.9(5)			& -0.10(5)		& 0.18(3)		& 3.18				& 5.35				\\
YbMnBi$_{2}$ \citep{Soh2019}		& 22.6(5)		& 7.8(5)			& -0.13(5)		& -0.37(4)		& 3.17				& 5.43				\\
\hline
BaMn$_{2}$Bi$_{2}$ \citep{Calder2014}	& 21.7(1.5)		& 7.85(1.4)		& 1.26(2)		& -0.87(15)		& 3.18				& 3.67				\\
BaMn$_2$As$_2$\citep{Ramazanoglu2017}& 40.5(2.0)		& 13.6(1.4)		& 1.8(3)			& -0.048(3)		& 2.95				& 3.36				\\
\hline \hline
\label{tab:conclusion}
\end{tabular}
\end{table*}

\subsection{First-Principles Calculations}

\begin{figure}
\includegraphics[width=3.5in]{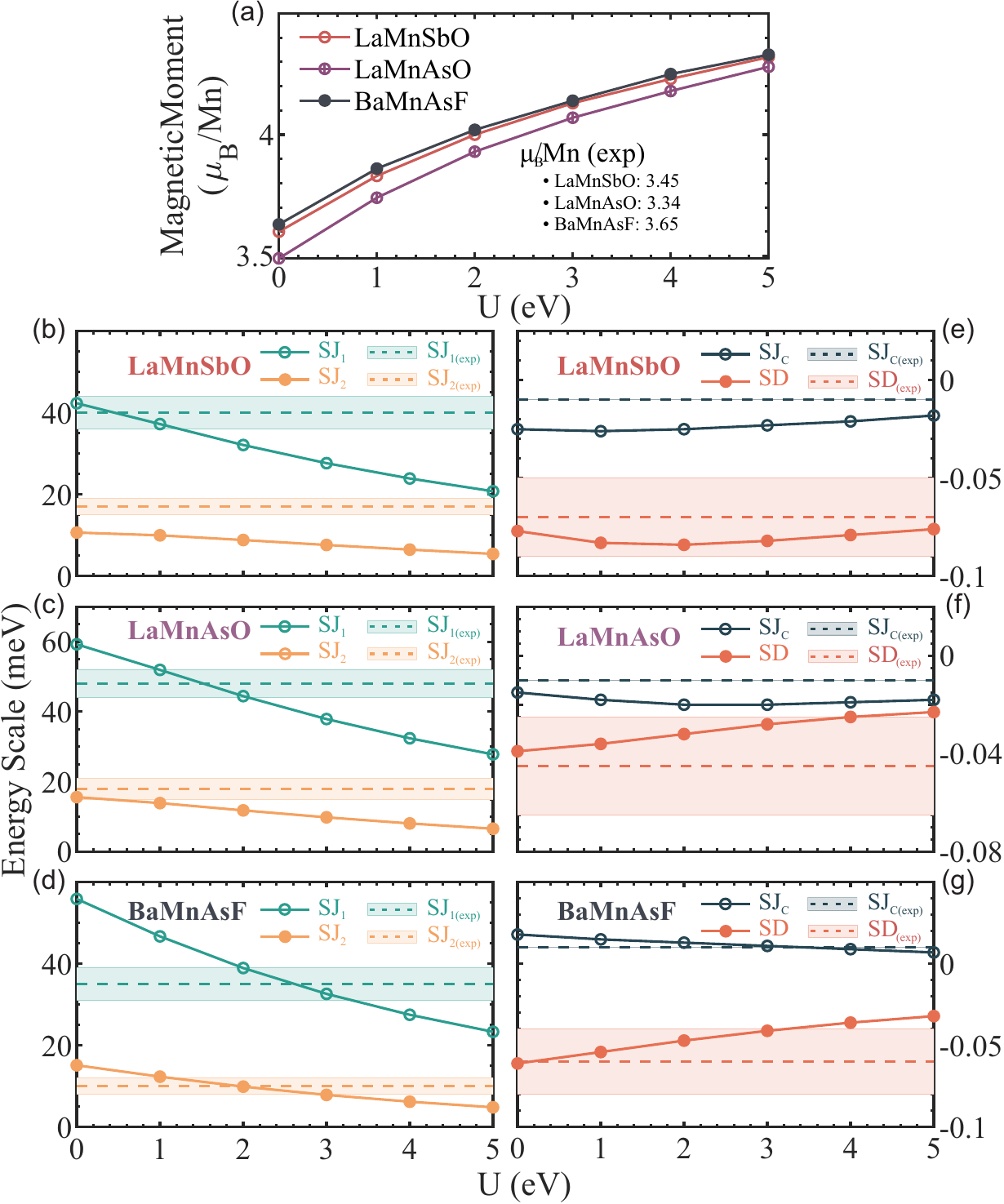}
\caption{(a) Magnetic moments localized on Mn for the AF1 state as a function of $U$ obtained from the DFT calculations. First principle calculations of (b-d) $SJ_1$, $SJ_2$, (e-g) $SJ_c$, and $SD$ vs experimentally determined values for LaMnSbO, LaMnAsO, and BaMnAsF respectively. Dashed lines are values obtained from the spin-waves analysis as listed in Table\ \ref{tab:params}. The shaded regions refer to experimental error found in $SJ_1$, $SJ_2$, $SJ_c$, and $D$.  Best agreement with theory and experiment occur at $U \simeq 0, 1$, and 2 eV for LaMnSbO, LaMnAsO, and BaMnAsF, respectively.}
\label{Fig:fpcalc}
\end{figure}

The magnetic ground states of LaMnSbO, LaMnAsO, and BaMnAsF, independent of  $U$, are correctly predicted using first-principles calculations.
Extracted $SJ_1$, $SJ_2$, $SJ_c$, and $SD$ for various $U$ values are shown in Fig.
\ref{Fig:fpcalc}.
As $U$ increases the localization of Mn d-states, $SJ_1$, $SJ_2$, and $SD$ values decrease in magnitude while $SJ_c$ experiences little change.
Quantitative agreement of theoretical $SJ_1$ and $SJ_2$ values are most consistent with INS experiments at $U\simeq 0$ eV for LaMnSbO, $U \simeq 1$ eV for LaMnAsO, and $U \simeq 2$ eV for BaMnAsF.
The C-type magnetic structure,found in LaMnSbO and LaMnAsO is readily explained by AFM intralayer and FM interlayer couplings, while the G-type magnetic structure of BaMnAsF arises from AFM interlayer coupling.

In agreement with experiment, DFT+$U$ calculations also find competing AF NN and NNN interactions within the square lattice layer.
For all compounds, experiments confirm that $2J_2/J_1<1$ which is a necessary condition for the observed intra-layer checkerboard AFM order.
However, it is somewhat surprising that this frustration is rather large.
For example, we find $2J_2/J_1=0.84$ for LaMnSbO which is responsible for low-lying M-point spin waves.
These results suggest two interesting possibilities.  The first is that the square lattice Mn pnictides may  adopt stripe AFM order for  larger $J_2$ values. Even more interesting is the possibility that such materials can be tuned into quantum disordered regime with $2J_2/J_1 \approx 1$ hosting a spin liquid\cite{Hong-Chen2012}.

Calculated Mn moments, as shown in Fig. \ref{Fig:fpcalc}, range from 3.49 to 4.33 $\mu_B/\text{Mn}$, increasing with $U$, showing greater localization as a function of increasing $U$, as expected. Additional electron-electron correlation, required to more accurately describe the INS data, slightly overestimates the on-site Mn moments found in these systems, i.e., 3.45 $\mu_{B,exp}$ $vs$ 3.60 $\mu_B$ for LaMnSbO at $U$ = 0 eV, 3.34 $\mu_{B,exp}$ $vs$ 3.74 $\mu_B$ for LaMnAsO at $U$ $\simeq 1$  eV, and 3.65 $\mu_{B,exp}$ $vs$ 4.02 $\mu_B$ for BaMnAsF at $U$ $\simeq  2$ eV \cite{McGuire2016,Zhang2016,Saparov2013}. Comparison of moment sizes and absolute values of $J_1$ and $J_2$ suggest small effective $U$ and  indicate a degree of delocalization of Mn $d-$electrons in all three systems.

\begin{figure}[!ht]
\includegraphics[width=2.5in]{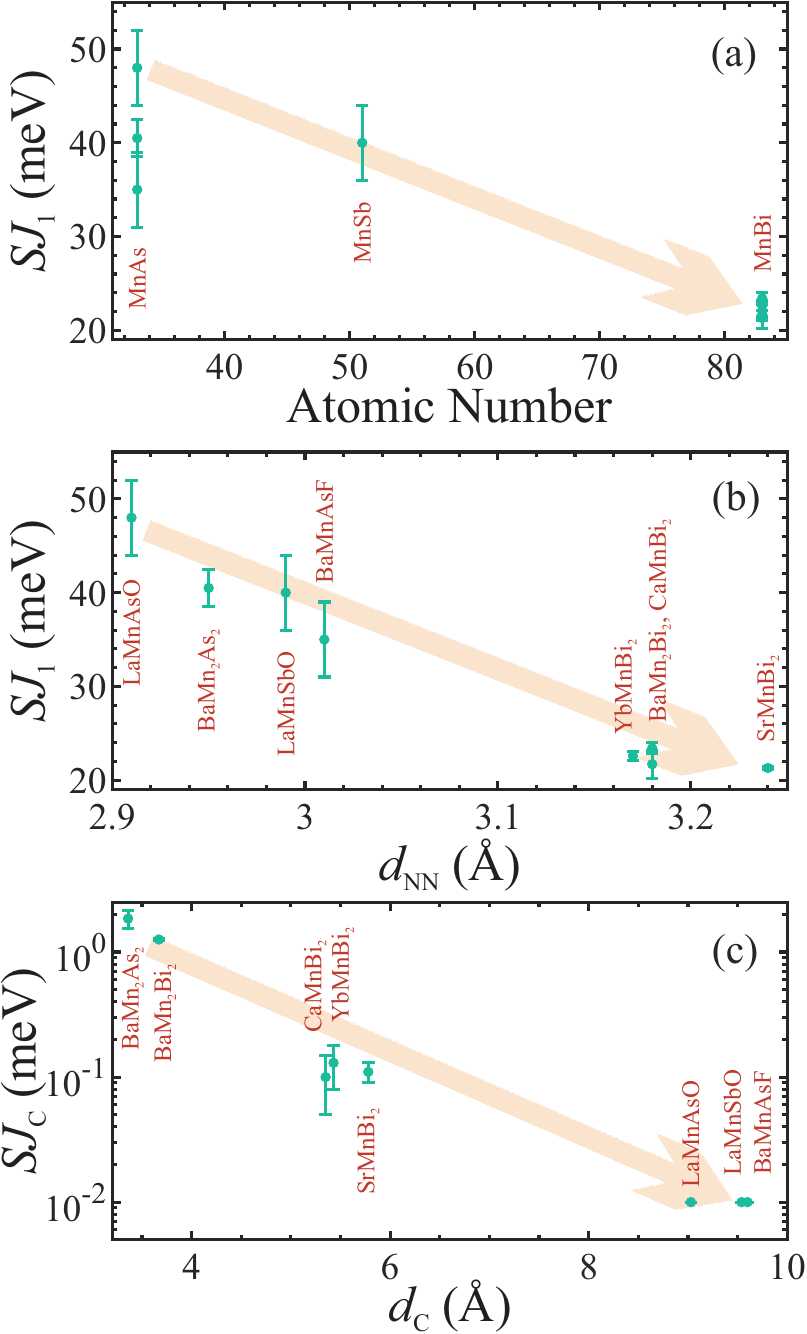}
\caption{Trends of (a) of $SJ_1$ on in-plane Mn-Mn NN distance $d_{\rm NN}$, (b) $SJ_1$ on various $Pn$ species, and (c) behavior of $SJ_c$ as a function of distance of adjacent Mn$Pn$ layers $d_{\rm c}$.  Graphs are based on data in Table\ \ref{tab:conclusion}}
\label{Fig:trends}
\end{figure}

\section{Conclusions}
We have  extracted the magnetic excitations of polycrystalline antiferromagnetic LaMnAsO, LaMnSbO, and BaMnAsF from inelastic neutron scattering data by removing signals from the sample holder, some phonons, and other background features.  We  analyzed the  magnetic spectra in the framework of inplane $J_1-J_2$ and out-of plane $J_c$ exchange coupling of a Heisenberg model using $\textsc{SpinW}$.  We also provide theoretical results using spin-polarized DFT + $U$ calculations, that to a large extent agree with the experimental results.  Our analysis shows that for all three samples $J_1$ and $J_2$ are antiferromagnetic  with a ratio $2J_2/J_1 < 1$  consistent with square lattice checkerboard order, but with $J_2$ large enough to consider effects of magnetic frustration.  We note that  the  largest  $2J_2/J_1$ ratio is obtained for LaMnSbO which may explain the  lower $T_N$ compared to the the other compounds\cite{Calder2014,Rahn2017,Soh2019,Ramazanoglu2017}.  The inter-plane coupling $J_c$ in all three systems is on order of $\sim 3\times10^{-4}J_1$ rendering these systems quasi-two-dimensional magnetic properties. Such a weak $J_c$ is due to the intervening rocksalt LaO and BaF layers, which effectively reduce the interlayer coupling compared to Mn-122 and Mn-112 square lattice antiferromagnets.

In Table\ \ref{tab:conclusion}, we list our results and published ones to show trends on the behavior of the exchange parameters in Mn$Pn$ layers. With regard to the intralayer exchange couplings, these are controlled by both steric effects from the rocksalt layers and the $Pn$ ligands (lattice parameters) and the hybridization of Mn with the specific $Pn$ ligand.  For example, in LaMn$Pn$O series, $J_1$ and $J_2$ are progressively reduced for heavier $Pn$ atoms due primarily to an increase in the Mn-Mn nearest-neighbor distance.  The larger distance between intralayer Mn atoms results in much weaker hybridization and exchange.
On the other hand, we can also compare LaMnSbO and BaMnAsF, which have nearly the same Mn-Mn distance but different $Pn$ ligands.  In this case, $SJ_1$ and $SJ_2$ are larger for the heavier Sb ligand, likely due to the increased hybridization from the extended p-orbitals of the Sb atom as compared to As. More generally,  Fig.\ \ref{Fig:trends}(a) shows that  $SJ_1$ decreases with the increase in in-plane  Mn-Mn NN distance, and Fig.\ \ref{Fig:trends}(b) shows  that  $SJ_1$ decreases as atomic number of $Pn$ gets larger.  In addition, Fig.\ \ref{Fig:trends}(c) shows significant decrease in $|J_c|$ as the spacing between adjacent Mn$Pn$ layers increases.

\subsection{Acknowledgments}
This research was supported by the U.S. Department of Energy, Office of Basic Energy Sciences, Division of Materials Sciences and Engineering.  Ames Laboratory is operated for the U.S. Department of Energy by Iowa State University under Contract No.~DE-AC02-07CH11358. This research used resources at the Spallation Neutron Source, a DOE Office of Science User Facility operated by the Oak Ridge National Laboratory

\bibliographystyle{apsrev4-2}

\bibliography{paper1111-a} 
\newpage
\newpage

{--------}
\newpage
\subsection{Supporting Information}
\setcounter{figure}{0}
\setcounter{equation}{0}
\setcounter{table}{0}

\renewcommand{\thefigure}{S\arabic{figure}}
\renewcommand{\theequation}{S\arabic{equation}}
\renewcommand{\thetable}{S\arabic{table}}
 
{\center{ 
 {\bf Spin dynamics in the antiferromagnetic oxy- and fluoro- pnictides: LaMnAsO, LaMnSbO, and BaMnAsF} \\

{\it Farhan Islam$^1$, Elijah Gordon$^1$, Pinaki Das$^1$,  Yong Liu$^1$, Liqin Ke$^1$, Douglas L. Abernathy$^2$, Rrobert J. McQueeney$^1$, and David Vaknin$^1$}\\
$^1$ {Ames Laboratory, Ames, Iowa 50011, USA} \\
$^2${Quantum Condensed Matter Division, Oak Ridge National Laboratory, Oak Ridge, Tennessee 37831, USA}
}
}

\author{Rrobert J. McQueeney}
\author{David Vaknin}

\subsection*{Cleaning up the INS data}
INS spectra were corrected for the background contribution from both aluminum (Al) phonons (coming from sample can) and from hydrogen (H) impurities due to inadvertent moisture adsorption on the polycrystalline surfaces from air exposure.

{\it Removing phonon signal near the magnetic scattering:}  By modeling the high Q trends of phonon intensities and extrapolating them to the low Q regions, phonon signals were subtracted from the magnetic region of interest. 

{\it Recoil scattering from hydrogen}:  The recoil scattering from H impurities is more prominent in the high-$Q$ regime, resulting in a broad feature with a $Q^2$ dependence extending beyond 100 meV. The contribution from the H scattering is estimated using a Gaussian function, $S_{\textrm{H}}({\textbf{Q}},E) = A_0 ~\textrm{exp}[(E - E_r)^2/\Delta^2]$, where $E_r$ is the H-recoil energy given by $E_r = \hbar^2Q^2/2m$, $\Delta$ is the energy width and $A_0$ is the scale factor. The Al sample holder contribution was measured with an empty Al can for each energy, and scaled and subtracted from the data. The spectra obtained after the above corrections were made are shown in Figs. \ref{Fig:ins150mev}.

\subsection*{Relation between $J_1$ and $J_2$ and their determination}
\begin{figure}[h]
\includegraphics[width=1.5in]{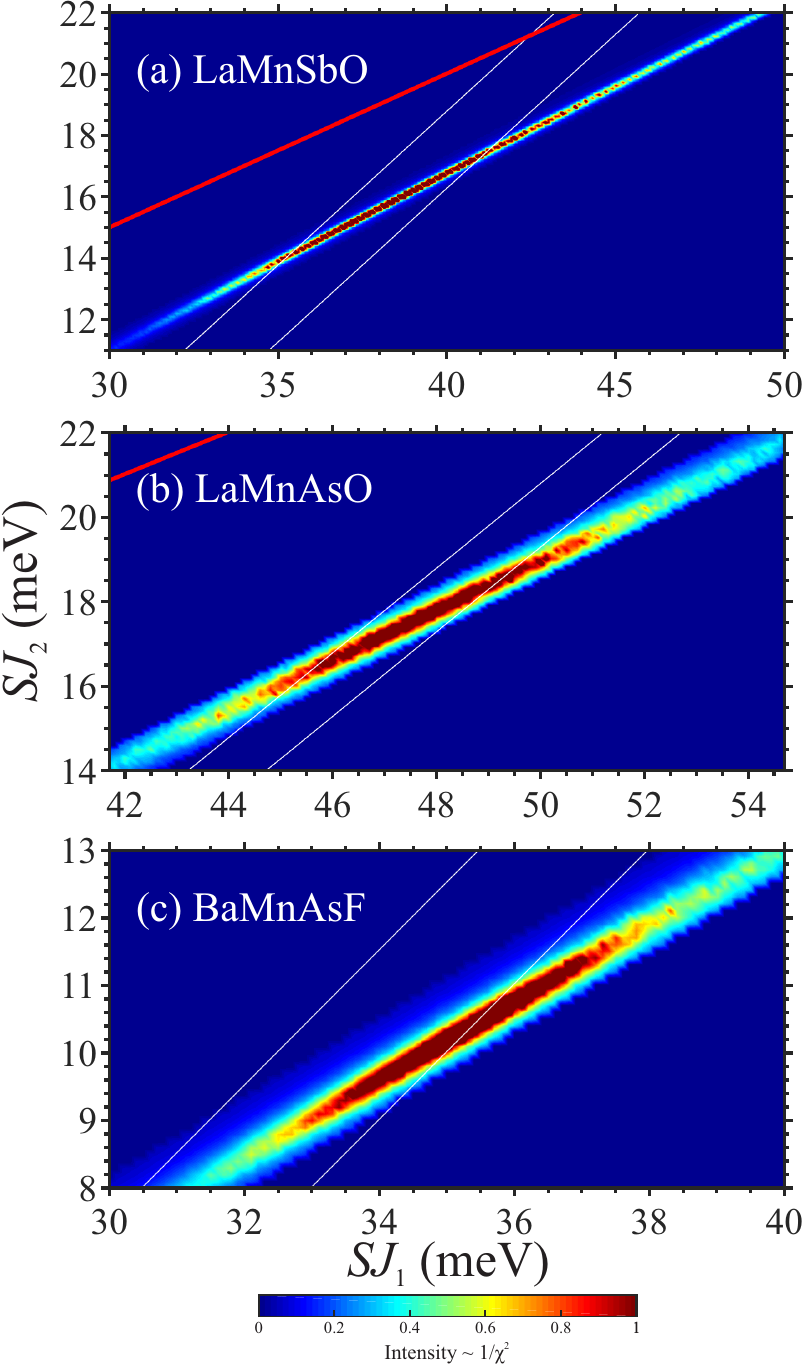}
\caption{3D plot of 1/$\chi^{2}(J_1,J_2)$  showing the relation between $J_1$ and $J_2$, obtained from best fits to energy cuts. The dashed line is extracted by determining the range of energy scale for the X-point and mapping it to corresponding $SJ_1$ and $SJ_2$ values. Solid red line is $SJ_1 = 2SJ_2$ above which the stripe structure is favored.}
\label{Fig:chi2fig2}
\end{figure}

Fixing $J_c$ and $D$, we systematically vary $J_1$ and $J_2$ to model  energy-cuts as shown in Fig.\ \ref{Fig:vanhovecut}. We calculate  $\chi^2$ values for numerous combinations  of $J_1$ and $J_2$ to search for its minimum to obtain the best fit to the data. A  color map of $\chi^{2}(J_1,J_2)$ is shown in Fig.\ \ref{Fig:chi2fig2} and the optimal values are listed in Table\ \ref{tab:params} (note that in Fig.\ \ref{Fig:chi2fig2}  we present 1/$\chi^{2}(J_1,J_2)$ for color enhancement purposes).  

\begin{figure}[h]
\includegraphics[width=2in]{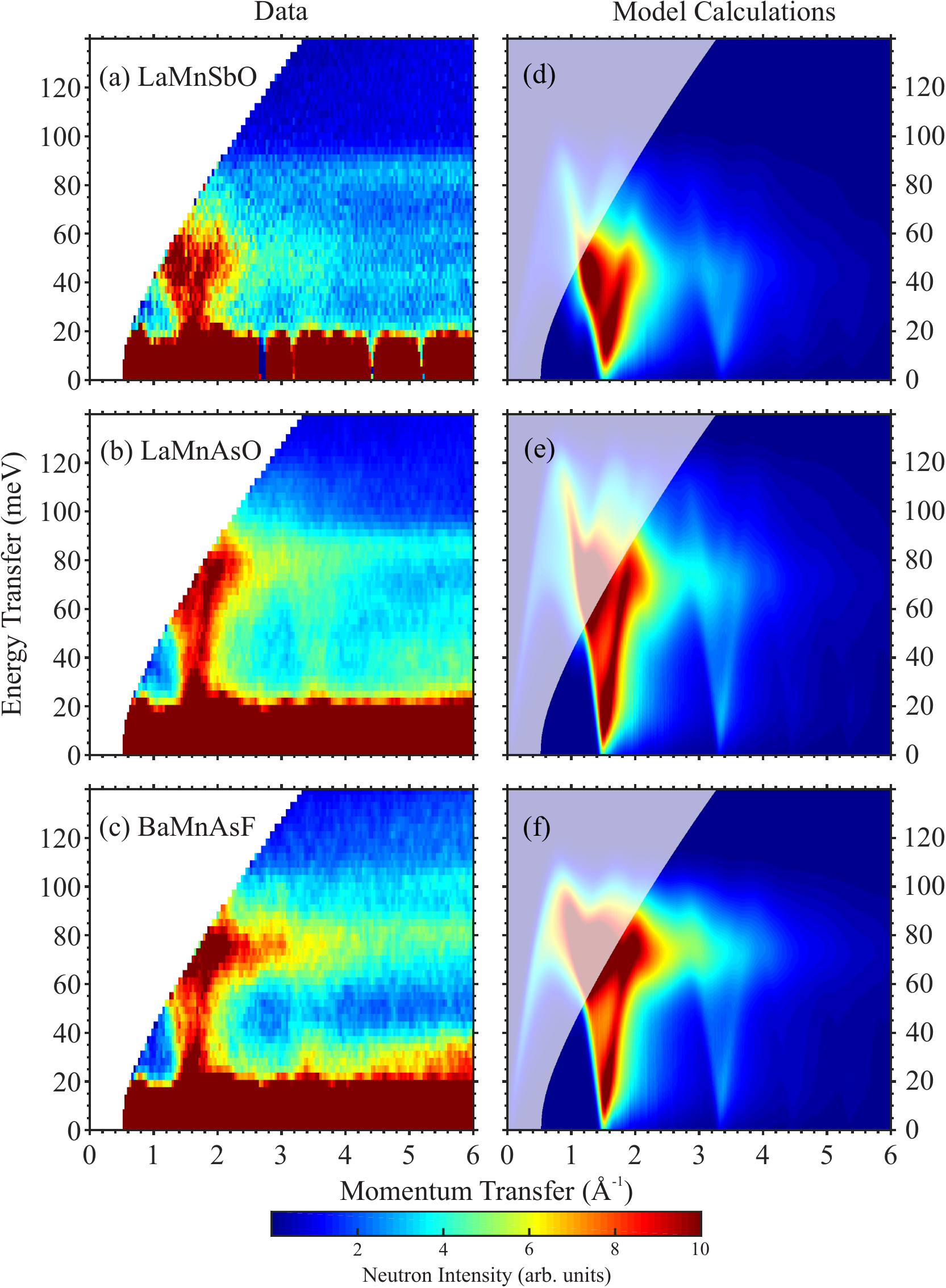}
\caption{(Left column) measured inelastic neutron scattering data at $E_i$ = 300 meV for (a) LaMnSbO, (b) LaMnAsO, and (c) BaMnAsF as indicated. (right column) (d-f) Corresponding calculated spectra using the best fit parameters given in  Table\ \ref{tab:params}. The shaded areas in the calculated panels are kinematically inaccessible regions for neutrons at the specified energy and set up.}
\label{Fig:ins300mev}
\end{figure}

We use the INS at $E_i = 300$ meV (Fig.\ \ref{Fig:ins300mev}) to estimate the spin-wave bandwidth corresponding to the X-point. 

\subsection*{Details of First-Principles Calculations}
\begin{figure}[h]
\includegraphics[width=1in]{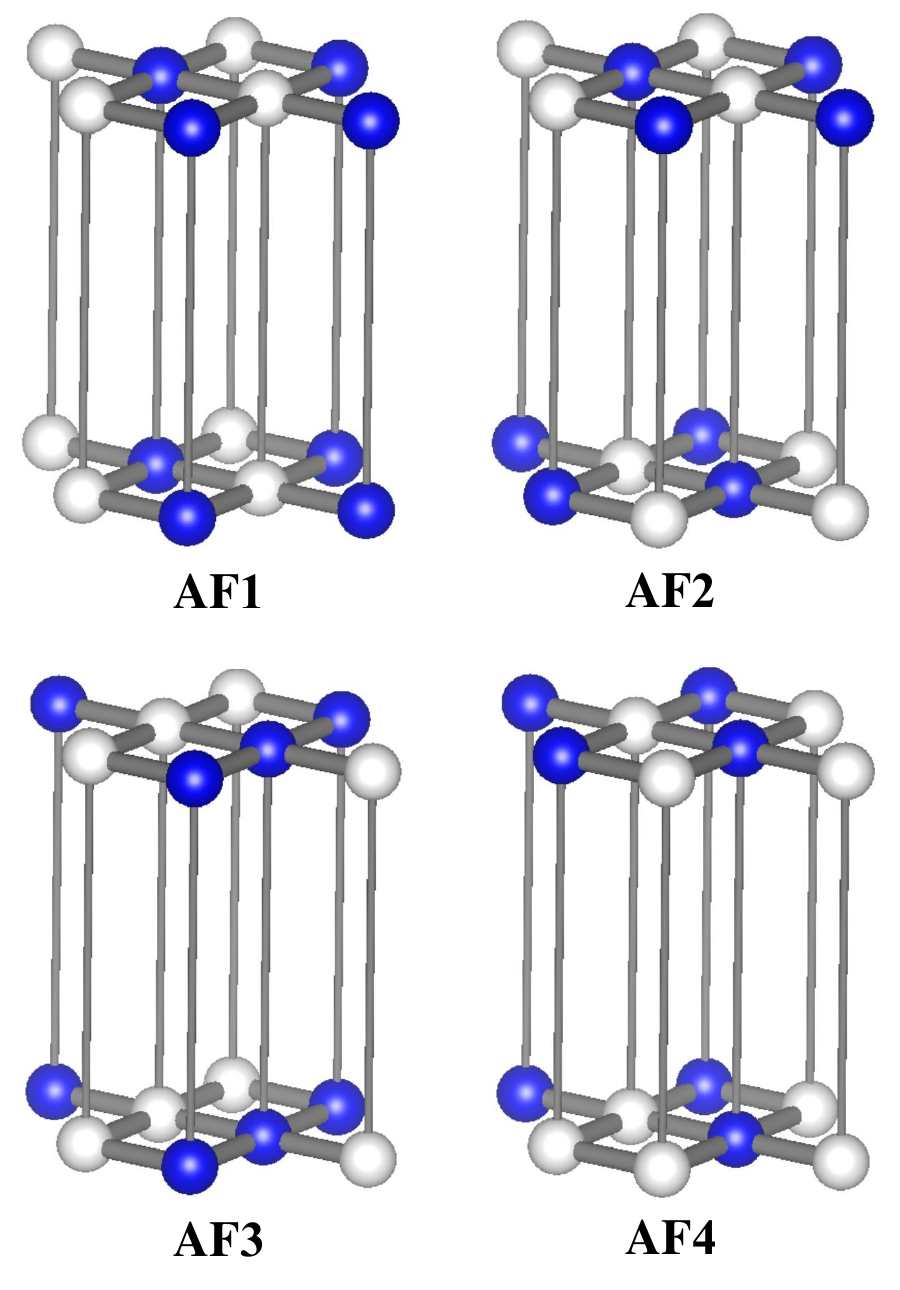}
\caption{$2a \times 2b \times 2c$ supercell with respect to the conventional unit cell used to construct construct states AF1 - AF4. These ordered spin states were used to extract $J_1$, $SJ_2$, and $SJ_c$. Blue (filled) spheres represent down-spin Mn and white (open) spheres represent up-spin Mn. Thick and thin bonds correspond to $J_1$ and $J_c$ interactions, respectively.}
\label{Fig:AFstructure}
\end{figure}

All isotropic spin exchange calculations used a $7 \times 7 \times 3$ $k$-point mesh within a $2a \times 2b\times 2c$ supercell (with respect to the conventional unit cell), MAE calculations for LaMnSbO and LaMnAsO used a $18 \times18\times 8$ k-point mesh within the conventional unit cell, and MAE calculations for BaMnAsF used a $18 \times 18 \times 4$ k-point mesh within a $1a\times 1b \times 2c$ supercell.
The total spin exchange energies, per supercell (16 f.u.), of the ordered spin states are given as
\begin{equation}
E_x=(n_1J_1+n_2J_2+n_3J_c)(S_\text{Mn})^2,
\end{equation}
where $E_x$ is the energy of different ordered spin state ($x = {\rm AF1} - {\rm AF4}$) relative to the ground state magnetic structure (AF1), and $S_\text{Mn}$ is the total spin, 5/2, localized on Mn$^{2+}$. The ordered spin states, values for $n_1-n_3$, and the relative energies of the ordered spin states are shown in Fig. \ref{Fig:AFstructure}, and Tables \ref{tab:sx1} and \ref{tab:sx2}.

\begin{table}[h]
\caption{Ordered spin states $n_1$-$n_2$ values used to extract $J_1$-$J_c$ using energy mapping analysis.}
\begin{tabular}{|c|c|c|c|}
\hline
Ordered Spin States & $n_1$ & $n_2$ & $n_3$ \\ \hline \hline
AF1                 & -32   & 32    & 16     \\ \hline
AF2                 & -32   & 32    & -16    \\ \hline
AF3                 & 0     & -32   & 16     \\ \hline
AF4                & -24   & 24    & 12     \\ \hline
\end{tabular}
\label{tab:sx1}
\end{table}

\begin{table}[h]
\caption{Relative energies (meV/16 f.u.) of the four ordered spin states used to calculate $SJ_1$-$SJ_c$ . Red
values signify the lowest energy state.}
\begin{tabular}{|c|c|c|c|c|c|}
\hline
                          &                          & \multicolumn{4}{c|}{Relative Energies, meV/16 f.u.}                           \\ \cline{3-6} 
\multirow{-2}{*}{System}  & \multirow{-2}{*}{$U$ (eV)} & AF1                         & AF2                          & AF3     & AF4    \\ \hline
                          & 0                        & {\color[HTML]{FE0000} 0.00} & 2.00                         & 1679.98 & 633.81 \\ \cline{2-6} 
                          & 1                        & {\color[HTML]{FE0000} 0.00} & 2.11                         & 1381.14 & 545.47 \\ \cline{2-6} 
                          & 2                        & {\color[HTML]{FE0000} 0.00} & 2.03                         & 1154.85 & 465.91 \\ \cline{2-6} 
                          & 3                        & {\color[HTML]{FE0000} 0.00} & 1.87                         & 991.27  & 400.91 \\ \cline{2-6} 
                          & 4                        & {\color[HTML]{FE0000} 0.00} & 1.67                         & 875.01  & 348.91 \\ \cline{2-6} 
\multirow{-6}{*}{LaMnSbO} & 5                        & {\color[HTML]{FE0000} 0.00} & 1.48                         & 792.71  & 307.27 \\ \hline
                          & 0                        & {\color[HTML]{FE0000} 0.00} & 1.17                         & 2249.12 & 873.94 \\ \cline{2-6} 
                          & 1                        & {\color[HTML]{FE0000} 0.00} & 1.48                         & 1934.91 & 761.54 \\ \cline{2-6} 
                          & 2                        & {\color[HTML]{FE0000} 0.00} & 1.60                         & 1670.39 & 653.50 \\ \cline{2-6} 
                          & 3                        & {\color[HTML]{FE0000} 0.00} & 1.60                         & 1462.89 & 562.08 \\ \cline{2-6} 
                          & 4                        & {\color[HTML]{FE0000} 0.00} & 1.53                         & 1302.24 & 486.94 \\ \cline{2-6} 
\multirow{-6}{*}{LaMnAsO} & 5                        & {\color[HTML]{FE0000} 0.00} & 1.42                         & 1177.00 & 425.39 \\ \hline
                          & 0                        & 0.00                        & {\color[HTML]{FE0000} -1.44} & 2046.78 & 814.73 \\ \cline{2-6} 
                          & 1                        & 0.00                        & {\color[HTML]{FE0000} -1.24} & 1762.11 & 687.49 \\ \cline{2-6} 
                          & 2                        & 0.00                        & {\color[HTML]{FE0000} -1.05} & 1531.27 & 581.03 \\ \cline{2-6} 
                          & 3                        & 0.00                        & {\color[HTML]{FE0000} -0.87} & 1349.55 & 495.11 \\ \cline{2-6} 
                          & 4                        & 0.00                        & {\color[HTML]{FE0000} -0.72} & 1205.99 & 425.94 \\ \cline{2-6} 
\multirow{-6}{*}{BaMnAsF} & 5                        & 0.00                        & {\color[HTML]{FE0000} -0.60} & 1090.78 & 369.89 \\ \hline
\end{tabular}
\label{tab:sx2}
\end{table}


\end{document}